\documentstyle[epsf]{article}
\begin{document}
\noindent{\em Preprint  KONS-RGKU-98-01}\hfill{\em Submitted to
Phys.Lett. A}
\vskip1cm

\begin{center}
{\Large \bf
Aharonov--Bohm scattering of neutral atoms \\ with induced electric dipole
moments}
\\[1cm]
{\bf  J\"urgen Audretsch$^{\dagger,}$\footnote{{\it e-mail:
Juergen.Audretsch@uni-konstanz.de}} and
Vladimir D. Skarzhinsky$^{\dagger, \ddagger,}$\footnote{{\it e-mail:
vdskarzh@sgi.lpi.msk.su}}}
\\[0.5cm]
$^\dagger$Fakult\"at f\"ur Physik der Universit\"at Konstanz,
Postfach M 673,\\ D-78457 Konstanz, Germany \\[0.5cm]
$^\ddagger$P.~N.~Lebedev Physical Institute, Leninsky prospect 53,
Moscow 117924, Russia
\\[1cm]

\begin{abstract}
\noindent
We investigate the scattering of neutral polarizable atoms from an
electrically charged wire placed in a homogeneous magnetic field.
The atoms carry an induced electric dipole. The reflecting wire is
discussed. We calculate the scattering amplitude and cross section for
the practically more important case that atoms
are totally absorbed at the surface of the wire. If the magnetic field
is present, there is a dominating Aharonov--Bohm peak in the forward
direction followed by decreasing oscillations for larger angles. An
experimental realization of this modulated Aharonov--Bohm scattering
should be possible.


\end{abstract}
\end{center}
\vspace{1cm}

Aharonov and Bohm have pointed out in their famous paper
\cite{Aharonov59} that electrons may interact quantum mechanically
with the vector potential $\vec{A}$ in such a way that they acquire
phase shifts when passing through regions where no classical
electromagnetic forces are acting on them. The Aharonov--Bohm (AB) phase is
a particular case of a geometrical phase \cite{Anandan97}. The effect has
been demonstrated experimentally \cite{Peshkin89}. For consequences
in the domain of quantum electrodynamics see \cite{Audretsch}. Aharonov
and Casher \cite{Casher84} showed that the interaction of the magnetic
moment of a quantum particle moving in the electric field of line
charge leads to an effect of the AB type.
A survey of interferometric observations of geometrical and
topological phases is given in \cite{Sterr97}.

Reversing the Aharonov--Casher situation, Wilkens \cite{Wilkens94}
showed that a corresponding quantum phase is to be expected for a
particle with an electric dipole moment moving in the radial field of
a not very realistic straight line of Dirac magnetic monopoles. A simple
experimental configuration much more practical to test the quantum phase
of electric dipoles moving in a combined magnetic and electric fields,
was proposed by Wei, Han and Wei \cite{Wei95}: The electric field of a
charged wire or of a cylindrical electrode polarizes a neutral atom. An
uniform magnetic field is applied parallel to the wire. Moving in these
fields atoms acquire a quantum phase. Below we will study the quantum
scattering of these atoms in detail.

The experimental set up is the following: a neutral atom (no permanent
electric dipole moment) with rest mass $M_0$ and electric polarizability
$\alpha$ is moving under the combined influence of electric and magnetic
fields. The electric field goes back to a charged straight wire lying
along the $z$-axes of a cylindrical coordinate system $(\rho, \varphi,
z).$ $\vec{E}=E \vec{e}_{\rho}$ with $E=\kappa/2\pi\rho.$ Superimposed is
a uniform magnetic field $\vec{B}=B \vec{e}_z$ pointing in the
$z$-direction. The moving atoms will be polarized by the electric field
in their rest frame which is equal to $\vec{E}+(1/c) \vec{v}\times \vec{B}$
with $\vec{v}$ being the velocity of the atom. The resulting Lagrangian is
\begin{equation}\label{L}
L = {1\over 2}Mv^2 + \alpha \,[\vec{B}\times\vec{E}]{\vec{v}\over c} +
{1\over 2}\alpha E^2 - U_0(\rho)\,.
\end{equation}
We have omitted higher order of $v^2/c^2$ and added a potential
$U_0(\rho)$ to allow for an interaction between atoms and the material
of the wire. The mass $M=M_0+\alpha B^2/c^2$ contains in addition to
the rest mass $M_0$ a very small magnetic mass.

The Lagrangian (\ref{L}) is equivalent to the Lagrangian of a charged
particle with mass $M$ and charge $e$ moving in the effective AB
vector potential $e\vec{A}(\rho) = (\hbar c\beta/\rho) \vec{e}_{\varphi}$
and in the electric scalar potential $U_E(\rho) =
- \alpha\kappa^2/(8\pi^2\rho^2).$ The magnetic field parameter $\beta$
is thereby $\beta =: \alpha \kappa B/(2\pi\hbar c)$ so that in our
case the vector potential $\vec{A}$ is proportional to the magnetic
field. The rotation of this effective potential vanishes so that it
yields no force on a classical particle but results in a topological
quantum phase (AB phase) for the quantum wave function. In
this sense it may be said that we are discussing among other effects
the AB scattering of an induced electric dipole.

In any case quantum scattering will not result in the pure AB
scattering because the electric field of the wire, or correspondingly
the influence of the scalar electric potential $U_E(\rho)$ may be essential.
The influence of $U_E(\rho)$ for the case of vanishing vector potential
$\vec{A},$ (no magnetic field present), has been studied by
Denschlag and Schmiedmayer \cite{Denschlag97} when investigating the
scattering of neutral polarizable atoms from a charged wire. They have
shown that the attractive potential $U_E(\rho)$ will make atoms fall
onto the wire unless the potential $U_0(\rho)$ prevents this or their
angular momenta are sufficiently large to pass by the wire and go to
infinity. It is the aim of this paper to work out analytically the
contributions of the different physical effects to the resulting
scattering and to provide expressions which can easily be evaluated
numerically.

We have to find a solution of the stationary Schr\"{o}dinger equation
corresponding to the Lagrangian (\ref{L}) (we omit the trivial
$z$-dependence of the wave function)
\begin{equation}\label{Se}
\left\{{\hbar^2\over 2M}\left[{\partial^2\over\partial\rho^2} + {1\over\rho}{\partial\over\partial\rho} - {1\over\rho^2}\left(-i{\partial\over \partial\varphi} - \beta\right)^2\right]  + {\alpha\kappa^2\over 8\pi^2\rho^2} - U_0(\rho) + {\cal E}\right\} \psi(\rho,\varphi) = 0
\end{equation}
which satisfies the scattering boundary condition at infinity
\begin{equation}\label{bc}
\psi (\vec{p}\,;\rho,\varphi)\sim {1\over 2\pi}\left[e^{ip\rho\cos(\varphi-\varphi_p)} + f(\varphi -\varphi_p)\; {e^{ip\rho}\over \sqrt{\rho}}\right]\,.
\end{equation}
The angle $\varphi_p$ fixes the direction of the incoming plane wave.
Below we will restrict to $\varphi_p=0.$

The scattering wave function can be obtained by the partial decomposition
\begin{equation}\label{Rs}
\psi(\rho, \varphi) = {1\over\sqrt{2\pi}}\sum_{m=-\infty}^{\infty} c_m R_m(\rho)\,e^{im\varphi}
\end{equation}
where the radial functions $R_m(\rho)$ satisfy the Bessel equation
\begin{equation}\label{Re}
R''_m (\rho) + {1\over\rho}\,R'_m(\rho) - {\nu^2\over \rho^2}\,R_m(\rho) - u_0(\rho)\,R_m(\rho) + p^2\,R_m(\rho) = 0\,.
\end{equation}
Here we have introduced
\begin{equation}\label{nu}
\nu^2 := (m-\beta)^2-{\alpha\kappa^2\over (2\pi\hbar)^2}(M_0+\alpha B^2/c^2) = m^2 - 2m\beta -\gamma^2,
\end{equation}
the wire parameter $\gamma$
\begin{equation}\label{gamma}
\gamma^2 := {\alpha\kappa^2\over (2\pi\hbar)^2}M_0\,,
\end{equation}
as well as
\begin{equation} \label{u00}
\hbar^2 u_0(\rho):=2M U_0(\rho) \quad {\rm and} \quad \hbar^2 p^2 := 2(M_0+\alpha B^2/c^2) {\cal E}.
\end{equation}

The scattering amplitude $f(\varphi)$ can be expressed in terms of phase
shifts $\delta_m$ which are defined by the asymptotic form of the radial
functions
\begin{equation} \label{asymR}
R_m(\rho) \sim \sqrt{2\over \pi p\rho} \cos\left(p\rho-{\pi m\over 2}-{\pi \over 4} + \delta_m\right)\,.
\end{equation}
Then the scattering amplitude reads
\begin{equation} \label{f}
f(\varphi) = {1\over\sqrt{2\pi}}\sum_{m=-\infty}^{\infty} f_m\,e^{im \varphi},
\quad f_m = {e^{-i{\pi\over 4}}\over\sqrt{p}}\left(S_m-1\right)\,, \quad
S_m = e^{2i\delta_m}\,.
\end{equation}

The parameter $\nu^2$ can take negative values for any values of the
parameters $\beta$ and $\gamma.$ This happens when $m$ is inside a
closed interval $[-m_-, m_+]$
\begin{equation}\label{m-+}
- m_- \leq m \leq m_+ \quad {\rm where} \quad m_{\pm} =
[\sqrt{\beta^2 + \gamma^2} \pm \beta]\,.
\end{equation}
$[a]$ denotes the integer part of $a.$ For these negative values of
$\nu^2$ the attractive potential $U(\rho)$ predominates over the
repulsive centrifugal potential and atoms can fall onto the wire
acquiring an infinite negative energy if it is a string with vanishing
radial extension at $\rho=0 \cite{Hagen96}.$

There are two ways to overcome this difficulty. One of them is to add
a repulsive hard core potential to the external fields
\begin{equation} \label{U0}
U_0 (\rho) = \left\{ \begin{array}{ll}
                     \infty & \mbox{if $\rho  < \rho_0$}  \\
                        0   &  \mbox{if $\rho > \rho_0$}
                     \end{array}    \right.
\end{equation}
representing a totally reflecting surface of the wire at $\rho=\rho_0.$
The second way is physically more realistic and reflects the actual
experimental situation: it is assumed that all polarized atoms which can
not escape but finally fall onto the wire are totally absorbed by the wire.
They do not take part in the scattering process furthermore. We turn now to
a discussion of these two cases.

\bigskip

{\bf 1. The totally reflecting wire}
\medskip

In the presence of a totally reflecting hard core potential (\ref{U0}) the wave function is equal to zero at $\rho=\rho_0$ and the solution of Eq.(\ref{Re}) has the form
\begin{equation} \label{R1}
R_m (\rho) = {1\over\sqrt{2\pi}} \left[J_{\nu}(p\rho) - \mu_m(a) H^{(1)}_{\nu}(p\rho)\right], \quad \mu_m(a) = {J_{\nu}(a)\over H^{(1)}_{\nu}(a)}, \quad a =: p\rho_0
\end{equation}
where $J_{\nu}(p\rho)$ and $H^{(1)}_{\nu}(p\rho)$ are the Bessel and the Hankel functions, correspondingly. It is easy to show that the coefficients $c_m$ of the decomposition (\ref{Rs}) which leads to the scattering wave function are equal to
\begin{equation} \label{c}
c_m = e^{i\pi m - i{\pi\over 2}\nu}.
\end{equation}
Then the partial scattering amplitudes read
\begin{equation} \label{f1}
f_m = {e^{-i{\pi\over 4}}\over\sqrt{p}}\left[e^{i\pi (m-\nu)} \left(1-2\mu_m(a)\right)-1\right].
\end{equation}
The hard core parameter $\mu_m(a)$ describes the additional scattering
caused by the screening potential $U_0.$ With (\ref{f1}) the problem can
essentially be solved.

We add that we obtain for the phase shifts $\delta_m$
\begin{equation} \label{S1}
S_m =: e^{2i\delta_m} = - e^{i\pi (m-\nu)}\,{H^{(2)}_{\nu}(a) \over H^{(1)}_{\nu}(a)}\,.
\end{equation}
Using the properties of the Hankel functions one can show that $|S_m|=1$
at positive as well as at negative values of $\nu^2,$ and that the value
of $S_m$ is independent of the sign of the imaginary part of $\nu$ in the
latter case. This means that the screening potential $U_0$ does indeed
prevent atoms from  penetrating the wire so that the scattering becomes
elastic in the presence of this potential.

The scattering amplitude and the cross section depend on the parameter
$a=p\rho_0.$ Therefore their behavior is quite different for low and
for high energies. The low energy scattering is more sensitive to the
AB effect since at high energies large values of $m$ dominate
the scattering amplitude for a given angle, and one can disregard the
AB parameter $\beta$ in partial scattering amplitudes. For
details we have to distinguish the cases of positive or negative values
of $\nu^2.$

For low energies, $(a = p\rho_0 << 1)$ and $\nu^2\geq 0$ the phase function is asymptotically equal to
\begin{equation} \label{S1ale}
S_m \approx e^{i\pi (m-\nu)}\,, \quad \delta_m = {\pi (m-\nu)\over 2}\quad \mbox{at} \quad m\geq m_+,\;m \leq m_-\,,
\end{equation}
and the hard core parameter
\begin{equation} \label{mule}
\mu_m(a) = {1\over 2}\left(1 - e^{-i\pi (m-\nu)}\,S_m\right) \approx 0
\end{equation}
goes to zero. These partial waves do not feel the totally reflecting potential. For $\nu^2<0$ the phase function can be written in the form \cite{Gradshteyn80}
\begin{equation} \label{S1ble}
S_m = e^{i\pi m}{I^{\ast}_{\nu}(a)\over I_{\nu}(a)}\,, \quad \mbox{at} \quad m_- < m < m_+
\end{equation}
where
\begin{equation} \label{I-+}
I_{\nu}(a) = \int_{-\infty}^{\infty}dt\, e^{i(a\cosh t + |\nu| t)} = 2\int_0^{\infty}dt\, e^{ia\cosh t}\;\cos |\nu|t\,.
\end{equation}
In this case the phase shifts
\begin{equation} \label{delta1le}
\delta_m = {\pi m \over 2}+\arctan \left[\tan\left(|\nu|\ln{a\over 2}\right)\, \tan{\pi |\nu|\over 2}\right]
\end{equation}
oscillate very quickly at $a\rightarrow 0$ since the low energy limit
$p\rightarrow 0$ is equivalent to the case of very thin wire
$\rho_0\rightarrow 0$ for which classical atoms will spiral into the center.
This leads to the situation where a small uncertainty of incoming atomic
beam energy will destroy the diffraction picture.

For high energies the method of partial waves becomes ineffective and we
need to use the WKB approximation to calculate the scattering cross section.
In the classical limit the hard core potential $U_0(\rho)$ and the scalar
potential $U(\rho)$ cause the scattering. The presence of the vector
potential $A(\rho)$ gives no effect.

It seems to be not very likely realistic anyway to find an experimental
realization for the scattering of atoms with an induced dipole moment
by a totally reflecting electrically charged wire. Moreover our
considerations show that the possibility to observe the AB
effect will depend very strongly on the experimental conditions. We
therefore turn now to the more promising situation of a totally absorbing
wire \cite{Leonhardt}.

\bigskip

{\bf 2. The totally absorbing wire}

\medskip

As pointed out above we assume that atoms which fall on the wire are
totally absorbed. They do not contribute to the stream of outgoing
particles. These inelastic collisions will happen for partial waves
with $m$ out of the interval $[-m_-, m_+]$ for which $\nu^2<0.$ As
usual total absorption can be taken into account by using vanishing
phase function $S_m=0$ for these inelastic channels.
\begin{equation}\label{S0}
S_m = 0 \quad \mbox{for} \quad -m_- \leq m \leq m_+ \,.
\end{equation}
The phase functions and phase shifts for the remaining modes with elastic
scattering are obtained from the radial functions
\begin{equation} \label{R2}
R_m (\rho) = {1\over\sqrt{2\pi}}\,J_{\nu}(p\rho)
\end{equation}
and are equal to
\begin{equation} \label{S2}
S_m = e^{i\pi (m-\nu)}, \quad \delta_m = {\pi\over 2}(m-\nu) \quad \mbox{for} \quad m \geq m_++1, \quad m \leq -m_--1 \,.
\end{equation}
We note that  $m_+\geq m_-.$

The elastic scattering amplitude $f(\varphi)$ allows us to
calculate the differential cross section for the elastic scattering
\begin{equation} \label{dcs}
{d\sigma^{\rm el}(\varphi)\over d\varphi} = |f(\varphi)|^2
\end{equation}
Taking into account the phase function (\ref{S0}) and (\ref{S2})
from (\ref{f}) we have
\begin{equation} \label{f2}
f(\varphi) = {e^{-i{\pi\over 4}}\over\sqrt{2\pi p}} \left[\Sigma_1(\beta, \varphi) + \Sigma_2(\beta, \varphi) + \Sigma_3(\beta, \varphi)\right]
\end{equation}
whereby
\begin{eqnarray} \label{Sigma}
\Sigma_1(\beta, \varphi) &:=& \sum_{m=m_++1}^{\infty}\left(e^{i\pi (m-\nu)}-1\right)\, e^{im\varphi}\,, \nonumber \\
\Sigma_2(\beta, \varphi) &:=& \sum_{m=m_-+1}^{\infty}\left(e^{i\pi (m-\tilde\nu)}-1\right)\, e^{-im\varphi}\,, \\
\Sigma_3 (\beta, \varphi) &:=& -\sum_{m=0}^{m_+}\, e^{im\varphi} -
\sum_{m=0}^{m_-}\, e^{-im\varphi} + 1 \nonumber
\end{eqnarray}
and $\tilde\nu = \nu(-\beta).$ The expressions (\ref{Sigma}) are invariant
under the substitution $\beta\rightarrow -\beta,\, \varphi\rightarrow -
\varphi.$ Therefore we can take $\beta > 0.$ We note that the scattering
amplitude is inverse proportional to $\sqrt{p}.$

It is impossible to evaluate $f(\varphi)$ analytically in a closed
expression since the parameter $\nu$ depends nonlinearly on $m.$ A
numerical calculation is difficult as well, because the series
(\ref{Sigma}) converge very slowly. However, the parameter $\nu$
reduces for large absolute values of $m$ to $\nu \approx |m-\beta|$
so that the partial waves approach asymptotically the partial waves
of the AB scattering amplitude
\begin{equation}\label{AB}
f_{AB}(\varphi) = {e^{-i{\pi\over 4}}\over\sqrt{2\pi p}}\,
\sum_{-\infty}^{\infty}\left(e^{i\pi (m-|m-\beta|)}-1\right)\,
e^{im\varphi} = - {e^{-i{\pi\over 4}}\over\sqrt{2\pi p}}\,
e^{i{1\over 2}\varphi}\,{\sin\pi\beta\over \sin {\varphi\over 2}}\,.
\end{equation}

It makes therefore sense to split $f(\varphi)$ in substracting from
the sums $\Sigma_1$ and $\Sigma_2$ (and only from these) the terms of
$f_{AB}(\varphi)$ with the same values of $m$ to obtain more rapidly
converging series. We combine these terms which are denoted by
$\Sigma_1^{AB}$ and $\Sigma_2^{AB}$ with the unmodified $\Sigma_3$ to
the total expression
\begin{equation}\label{dab1}
f_{AB}^{\rm mod}(\varphi) = {e^{-i{\pi\over 4}}\over\sqrt{2\pi p}}
\left[\Sigma_1^{AB}(\beta, \varphi) + \Sigma_2^{AB}(\beta, \varphi) +
\Sigma_3(\beta, \varphi)\right]
\end{equation}
with
\begin{eqnarray} \label{Sab}
\Sigma_1^{AB}(\beta, \varphi) &=& \sum_{m=m_++1}^{\infty}\left(e^{i\pi (m-|m-\beta|)}-1\right)\, e^{im\varphi}\,, \nonumber \\
\Sigma_2^{AB}(\beta, \varphi) &=& \sum_{m=m_-+1}^{\infty}\left(e^{i\pi (m-|m+\beta|)}-1\right)\, e^{-im\varphi}\,.
\end{eqnarray}
$f_{AB}^{\rm mod}(\varphi)$ has a simple physical interpretation. By
construction it is the modified AB amplitude which is obtained by taking
into account the absorption of the respective partial waves. This splitting
leads for the scattering amplitude $f(\varphi)$ to
\begin{equation} \label{f3}
f(\varphi) := f_{AB}^{\rm mod} (\varphi) + f_w(\varphi)
\end{equation}
with
\begin{equation}\label{w}
f_w (\varphi) = {e^{-i{\pi\over 4}}\over\sqrt{2\pi p}}
\left[\delta\Sigma_1(\beta, \varphi) + \delta\Sigma_2(\beta, \varphi)\right]
\end{equation}
whereby
\begin{eqnarray} \label{dSig}
\delta\Sigma_1(\beta, \varphi) &=& \sum_{m=m_++1}^{\infty}\left(e^{i\pi (m-\nu)}-e^{i\pi(m - |m-\beta|)}\right)\, e^{im\varphi} \nonumber \\
&=& 2i e^{i\pi\beta} \sum_{m=m_++1}^{\infty} e^{i{\pi\over 2}{\beta^2+\gamma^2\over m-\beta+\nu}}\,\sin{\pi\over 2}{\beta^2+\gamma^2\over m-\beta+\nu}\, e^{im\varphi}\,, \\
\delta\Sigma_2(\beta, \varphi) &=& \sum_{m=m_-+1}^{\infty}\left(e^{i\pi (m-\tilde\nu)}-e^{i\pi(m - |m+\beta|)}\right)\, e^{-im\varphi}  \nonumber  \\
&=& 2i e^{-i\pi\beta} \sum_{m=m_-+1}^{\infty} e^{i{\pi\over 2}{\beta^2+\gamma^2\over m+\beta+\tilde\nu}}\,\sin{\pi\over 2}{\beta^2+\gamma^2\over m+\beta+\tilde\nu}\, e^{-im\varphi}\,.\nonumber
\end{eqnarray}
Details of the intermediate steps of the calculation are omitted.

The modified AB amplitude (\ref{dab1}) can be easily reformulated:
\begin{equation}\label{dab2}
f_{AB}^{\rm mod}(\varphi) = - {e^{-i{\pi\over 4}}\over\sqrt{2\pi p}}\, e^{i{m_+-m_- \over 2}\varphi}\,{1\over \sin {\varphi\over 2}}\,\sin\left(\pi\beta + {m_++m_-+1\over 2}\varphi \right)\,.
\end{equation}
It turns out to be the well known AB amplitude $f_{AB}(\varphi)$ with
modified $\sin\pi\beta$ factor. We have now obtained the scattering
amplitude $f(\varphi)$ as a sum of a closed analytical expression and two
quickly converging series. We are therefore prepared for a numerical
evaluation as well as a physical interpretation of the different
contributions to $f(\varphi).$

If there is a completely absorbing wire whose radius $\rho_0$ is large
as compared to the wavelength of atoms $p\rho_0 >> 1,$ then all atoms
which are incident with an impact parameter $a = (m-\beta)/p < \rho_0$,
or with angular momentum $|m-\beta| < p\rho_0$, will be absorbed by
the wire. Accordingly we have to replace in this case the $m_{\pm}$ in
(\ref{Sigma}) by the integer part of $p\rho_0.$

Eqs.(\ref{dSig}) and (\ref{dab2}) show that because the absorbing wire
cuts out modes, all parts of the scattering amplitude have an oscillatory
behaviour. Also the modified AB amplitude is modulated in
this sense. In addition, $f_{AB}^{\rm mod}(\varphi)$ and $f_w(\varphi)$ both
diverge for forward scattering $\varphi=0,$ although $f_w(\varphi)$
diverges slowlier than $f_{AB}^{\rm mod}(\varphi).$ It is therefore the
region of small angles around the forward direction into which the atoms
are scattered predominantly and where experimental tests should be
located. We turn to this region now.

The most singular term in the forward direction arises from
$f_{AB}^{\rm mod}(\varphi)$
\begin{equation}\label{dab3}
f_{AB}^{\rm mod}(\varphi<<1) = - {e^{-i{\pi\over 4}}\over\sqrt{2\pi p}}\,
e^{i{m_+-m_- \over 2}\varphi}\,{\sin\pi\beta\over \sin {\varphi\over 2}}\,
\cos{m_++m_-+1\over 2}\varphi\,.
\end{equation}
The oscillating factor modulates the AB amplitude so that
narrow peaks and deeps of the width $\delta\varphi=\pi/2\gamma$ arise
which are located at $\varphi=n\pi/\gamma$ and $\varphi=(n+1/2)\pi/\gamma,$
correspondingly.

The singularity of $f_w(\varphi)$ for $\varphi=0$ is connected with large
value of $m.$ Neglecting small terms in Eq.(\ref{dSig}) and using the
formula 5.4.3(2) in \cite{Prud} we obtain
\begin{eqnarray}\label{s}
\sum_{m=0}^{\infty}{1\over m+\alpha}\,e^{im\varphi} &=& \beta(\alpha)\,
e^{i\alpha (\pi - \varphi)} \nonumber \\
&+& {1\over 2}\int_{\varphi}^{\pi} e^{i[-\alpha\varphi +
(\alpha - 1/2) t]}{dt\over \sin (t/2)}
\end{eqnarray}
whereby
$$
\beta(\alpha) = {1\over 2} \left[\psi\left({\alpha +1 \over 2}\right) -
\psi\left({\alpha \over 2}\right)\right]
$$
and $\psi(\alpha)$ is the psi - function. This results in
\begin{equation}\label{sds}
f_w(\varphi << 1) \sim (\beta^2 + \gamma^2) \cos \pi\beta \ln\varphi\,.
\end{equation}

If one neglects the magnetic mass $\alpha B^2/c^2 << M_0,$ what will
usually be justified, the scattering amplitude depends periodically on
the magnetic field parameter $\beta.$ This follows from the gauge
invariance of the wave equation (\ref{Se}) in this case. We may therefore
consider this parameter to be restricted to $0\leq\beta\leq1.$ If one is
interested to study how the influence of the magnetic field modifies
the situation where only the charged wire is present, it is useful to
compare the case $\beta=0$ (no AB contribution) with the
case $\beta=1/2$ (maximal AB effect).

We may assume that the electric field parameter $\gamma$ is larger than
the restricted $\beta.$ In this case $\beta$ plays no role in
$f_w(\varphi)$ for small angles and the switching on of the magnetic
field results solely in the appearance of the modulated AB
term $f_{AB}^{\rm mod}.$ In addition we have that the width of the
oscillation is about $\pi/2\gamma.$ If one want to resolve these
oscillations one should choose a not too big $\gamma.$ In any case
we have for only small angles around the forward direction
$$
{f_{AB}^{\rm mod}\over f_w} \sim (\varphi \ln\varphi)^{-1}
$$
which is diverging for $\varphi=0,$ so that the appearance the
AB term for the non-vanishing magnetic field should be
observable in any case. If the angular resolution is good enough one
can also observe the modulation of the AB cross section. If not,
the cross section will be averaged and we obtain half of the
AB result because the average of $\cos^2x$ is 1/2.

We turn to two realizations of the parameters of the experimental set up.
One realization is discussed in \cite{Wei95}. The value of $\beta =
\alpha \kappa B/(2\pi\hbar c) = 0.5$ can be obtained for alkali atoms
with the polarizability $\alpha \sim 10\times 10^{-40} {\rm F m^2}$
placed in an electric field $E \sim 10^7 {\rm V/m}$ of the wire of the
radius $\sim 1 {\rm mm}$ (this corresponds to $\kappa \sim 10^4 {\rm V}$)
and in a magnetic field $B \sim 5 {\rm T}.$  In this case we have
$\beta^2/\gamma^2 = \alpha B^2/M_0 c^2 \leq 10^{-11}.$ Obviously we can
neglect $\beta$ as compared to $\gamma.$ In this case the distances
between two peaks are extremely small so that the cross section will
get the factor 1/2. An estimation of $\gamma$ of the order of 10 was
given in \cite{Denschlag97}. In this case it should be possible to
separate the peaks.

The figures show for two values of the wire parameter $\gamma$ the
differential cross section for small angles. The oscillations go back
to the absorption of particles on the wire and are also present for
vanishing magnetic field ($\beta=0,$ thin lines). Switching on the
magnetic field ($\beta=1/2$, bold lines) leads to a shift of the
oscillations. What has not been shown is that for smaller angles
$\varphi$ the cross section with magnetic field is considerably larger
for both values of $\gamma.$ This goes back to the fact that the AB
scattering dominates for very small angles.

\bigskip

{\bf Acknowledgments}
\medskip

J.A. gratefully acknowledges stimulating discussion with B.Finck von
Finckenstein, U.Leonhardt and M.Wilkens.
V.~S.~thanks Prof.J.~Audretsch and the members of his group for the
friendly atmosphere and the hospitality at the University of Konstanz.
This work was supported by the Deutsche Forschungsgemeinschaft.
\newpage
{\begin{figure}[h]
\let\picnaturalsize=N
\def\picsize{3.2in}
\def\picfilename{c.ps}
\ifx\nopictures Y\else{\ifx\epsfloaded Y\else\input epsf \fi
\let\epsfloaded=Y
\centerline{\ifx\picnaturalsize N\epsfxsize
\picsize\fi
\epsfbox{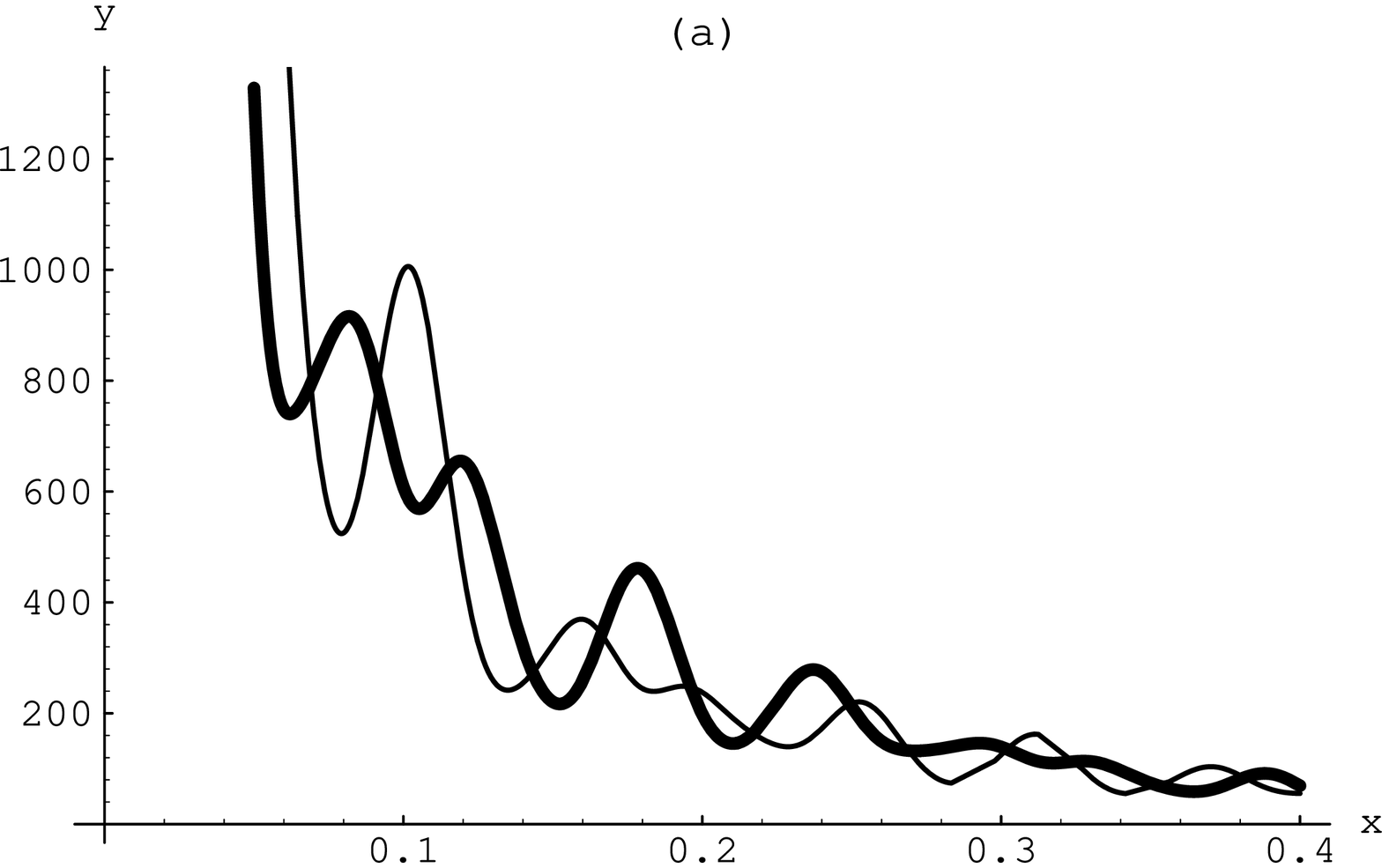}}}\fi
\let\picnaturalsize=N
\def\picsize{3.2in}
\def\picfilename{d.ps}
\ifx\nopictures Y\else{\ifx\epsfloaded Y\else\input epsf \fi
\let\epsfloaded=Y
\centerline{\ifx\picnaturalsize N\epsfxsize
\picsize\fi
\epsfbox{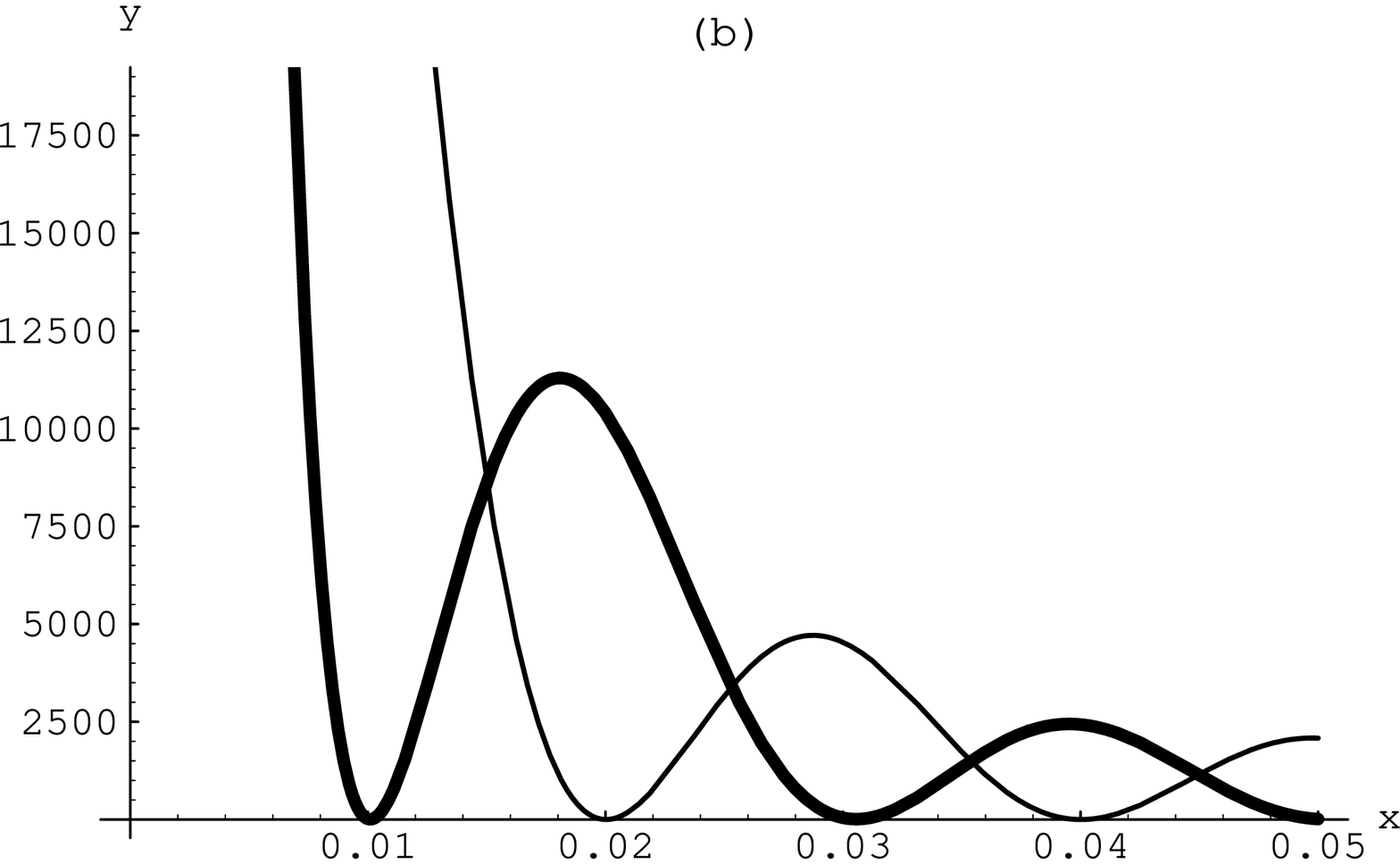}}}\fi
\caption{
Differential scattering cross section ($x=\varphi, \quad
y=2\pi p |f(x)|^2$) for the case of a totally absorbing wire
and vanishing magnetic field ($\beta=0$, thin lines) or non vanishing
magnetic field ($\beta=1/2$, bold lines).The wire parameter
$\gamma=5.1$ in Figure (a) and $\gamma=50.1$ in Figure (b).}
\end{figure}
}
\newpage

\end{document}